\documentclass[aip,jcp,twocolumn,10pt,amsmath,amssymb,color,epsfig,graphicx,bm]{revtex4-1}
\usepackage{graphicx}
\usepackage{dcolumn}
\usepackage{bm}
\usepackage{subfigure}

\begin{document}

\title{Crystal structure prediction using the Minima Hopping method}

\author{Maximilian Amsler}
\email{M.Amsler@unibas.ch}
\affiliation{Department of Physics, Universit\"{a}t Basel,
Klingelbergstr. 82, 4056 Basel, Switzerland \\}
\author{Stefan Goedecker}
\email{Stefan.Goedecker@unibas.ch}
\affiliation{Department of Physics, Universit\"{a}t Basel,
Klingelbergstr. 82, 4056 Basel, Switzerland \\}

\date{\today}

\begin{abstract}
A structure prediction method is presented based on the Minima Hopping method. Optimized moves on the configurational enthalpy surface are performed to escape local minima using variable cell shape molecular dynamics by aligning the initial atomic and cell velocities to low curvature directions of the current minimum. The method is applied to both silicon crystals and binary Lennard-Jones mixtures and the results are compared to previous investigations. It is shown that a high success rate is achieved and a reliable prediction of unknown ground state structures is possible.

\end{abstract}


\maketitle

\section{Introduction}
Predicting structures of periodic systems~\cite{oganov_2010} is one of the most challenging tasks in material sciences. Not only in material design under various external conditions, but also in biology and pharmacy there is an increasing demand for efficient and reliable structure prediction methods.~\cite{neumann_2008,day_2009} A very common way of predicting favorable structures is extracting known structures from databases of structures previously found in similar materials. The energetically most stable structure is identified and gives the putative ground state. However, this approach has a limited success rate when the ground state is an unknown structure, which can only be found by performing an extensive search. Similarly, data mining is capable of predicting new crystalline structures based on a huge set of experimental data and/or \textit{ab initio} calculations.~\cite{Curtarolo2003,fischer2006}

Recently, more advanced methods for crystal structure prediction have been developed and applied, which allow a systematic search for the ground state structure based solely on the system's composition and the external conditions. The most promising of these methods and their applications on crystal structure prediction include simulated annealing,~\cite{salamon2002,deem_1989,pannetie_1990,boisen_1994,schoen_1996} genetic algorithms~\cite{goldberg_1989,bush_1995,woodley_1999,bazterra_2002,woodley_2004,abraham_2006,glass_2006,oganov_2006,trimarchi_2007} and metadynamics.~\cite{laio_2002,martonak_2003,martonak_2005,martonak_2006,behler_2008} Some of the methods, their advantages and drawbacks are characterized and discussed by Oganov \textit{et al.},~\cite{oganov_2008} giving a good overview over the aforementioned structure prediction methods.

The Minima Hopping (MH) method is an algorithm which allows an efficient exploration of a high dimensional potential energy surface of complex systems, while progressing toward the global minimum structure.~\cite{goedecker_2004} It has been successfully applied to various isolated systems such as Lennard-Jones and silicon clusters,~\cite{goedecker_2004,goedecker_2005,hellmann_2007} doped silicon fullerenes,~\cite{willand_2010} complex biological molecules~\cite{roy_2009} and large gold clusters.~\cite{bao_2009} 
The MH method has also been used in other applications, such as providing realistic atomic force microscopy (AFM) tips for AFM simulations.~\cite{ghasemi_2008,pou_2009,amsler_2009} In this paper we present an approach for structure prediction of crystals by generalizing the MH method to periodic systems with pressure constraint. The method was tested on two benchmark systems, silicon crystals and binary Lennard-Jones (BLJ) mixtures. Finding the ground state of the Si$_{64}$ cell presents a challenging task due to its multi-funnel character. Volume constraints are used to demonstrate the capability of predicting porous silicon crystal structures. \textit{Softening}, a procedure to modify initial molecular dynamics velocities, is used to increase the efficiency. To demonstrate the effect of \textit{softening} statistical data are collected for its application on a BLJ mixture in a small cell. Furthermore, a larger, much-studied BLJ mixture is investigated and a new putative ground state structure is presented.

\section{Method and Application}
\subsection{Generalizing the Minima Hopping to periodic systems}

The initial implementation of the MH method was capable of performing a search for low-lying minima on the \mbox{$3N$-dimensional} potential energy surface \mbox{$E=E(\mathbf{r}_i)$, $i=1, \dots ,N$}, of an isolated molecule or a periodic system in a rigid box with $N$ atoms. Starting from some current local minimum on the potential energy surface, an escape step is performed using a short molecular dynamics (MD) simulation which is stopped as soon as $md_{min}$ potential energy minima have been found along the MD trajectory. Then, a local geometry relaxation is performed. The new minimum is either accepted or rejected, depending on its energy and a threshold value \mbox{$E_{diff}$}, which is updated by a feedback such that half of the new configurations are accepted. To avoid exploring previously visited regions of the potential energy surface, the kinetic energy $E_{kin}$ is increased when already known structures are revisited. Then, the cycle is repeated starting with a new MD escape.

However, it is essential for the purpose of structure prediction in periodic systems that not only the atomic positions are allowed to be optimized, but also the cell shape, especially when external constraints are imposed. Furthermore, using a variable cell shape reduces the enthalpy barriers for phase transitions. Hence, to generalize the MH method for periodic systems with variable cell shape, the degrees of freedom are augmented by the three variable cell vectors $\mathbf{a}$, $\mathbf{b}$ and $\mathbf{c}$, describing the edges of the simulation cell. They are combined to a $3\times3$ tensor \mbox{$h=\{\mathbf{a},\mathbf{b},\mathbf{c}\}$}. The atomic positions can be expressed by vectors in lattice coordinates $\mathbf{s}_i$ according to \mbox{$\mathbf{r}_i=h\mathbf{s}_i$}. The potential energy surface \mbox{$E=E(\mathbf{s}_i,h)$} is now a function of the cell vectors and the atomic positions with respect to the cell vectors. When an external pressure $P$ is applied, the potential energy is replaced by the configurational enthalpy \mbox{$H_c=E(\mathbf{s}_i,h)+P\Omega(h)$}, where $\Omega=\det(h)$ is the volume of the simulation cell. Periodic boundary conditions are applied for the evaluation of $H_c$.

Finding the local extrema of the configurational enthalpy is equivalent to finding the set of coordinates for which the following conditions are satisfied:
\begin{align}
\frac{\partial H_c}{\partial s_i^\gamma}&=\frac{\partial E}{\partial s_i^\gamma}=0\\  
\frac{\partial H_c}{\partial h_{\alpha\beta}}&=\frac{\partial E}{\partial h_{\alpha\beta} }+P\frac{\partial \Omega}{\partial h_{\alpha\beta}}=0
\end{align}
where $s_i^\gamma$, $\gamma \in \{x,y,z\}$ are the components of the fractional coordinate  of atom $i$ and $h_{\alpha\beta}$ denote the elements of the tensor $h$.

Similarly, for the escape step, the MD needs to be performed taking into account the additional cell parameters. Hence, both the atomic positions in lattice coordinates $\mathbf{s}_i(t)$ and the lattice vectors $h(t)$ are time-dependent. A Lagrangian to perform variable cell shape MD at constant pressure $P$ was proposed by Parrinello and Rahman in 1980~\cite{parrinello_1980} and has been widely applied: 
\begin{equation}
L=\sum_{i=1}^N \frac{m_i}{2}\dot{\mathbf{s}}_i^T g \dot{\mathbf{s}}_i -
\sum_{i=1}^N \sum_{i<j}^N \phi(r_{ij})+
\frac{W}{2}\mathrm{Tr}(\dot{h}^T\dot{h})-P\Omega
\label{eq:lagrange_w} 
\end{equation}
where \mbox{$g=h^Th$} is the metric tensor, $W$ is the fictitious mass of the simulation cell and $m_i$ is the mass of atom $i$. The first two terms in equation \eqref{eq:lagrange_w} correspond to the kinetic and potential energy of the atoms, respectively. The third and the fourth term correspond to the kinetic and potential energy of the simulation cell, respectively. The interaction between the atoms are described by a radial pair potential \mbox{$\phi(r_{ij})$} where \mbox{$r_{ij}=\left|\mathbf{r}_i-\mathbf{r}_j\right|$}. However, the formalism can readily be extended to any many-body potential which are expressed by atomic positions.

The symmetrized stress tensor $\Pi$ can be written as 
\begin{equation}
\Pi=\frac{1}{\Omega}\left(\sum_{i=1}^{N}m_i\mathbf{v}_i\mathbf{v}_i^T+ f^h h^T\right)
\label{eq:stress}
\end{equation}
where we use the cell gradients of the potential energy \mbox{$f^h_{\alpha \beta}=\frac{\partial E(\mathbf{s}_i,h)}{\partial{h_{\alpha,\beta}}}$} and define \mbox{$\mathbf{v}_i=h\dot{\mathbf{s}_i}$}. The second term in equation \eqref{eq:stress} is related to the components of the negative volume-normalized strain derivatives of the energy \mbox{$\sigma_{\alpha\beta}=-\frac{1}{\Omega}\frac{\partial E}{\partial \epsilon_{\alpha\beta}}$} (the stress tensor for the static case) by \mbox{$f^h h^T=-\Omega\sigma$}, where $\epsilon$ is the strain tensor (see Ref.~[\onlinecite{bucko_2005}]).

Using the Lagrange's equation, the equations of motion for the atomic positions in lattice coordinates and the cell vectors are then given by
\begin{align}
\ddot{\mathbf{s}}_i&=-\frac{1}{m_i}\sum_{i\neq j}\frac{1}{r_{ij}}\frac{\partial \phi(r_{ij})}{\partial{r_{ij}}}(\mathbf{s}_i-\mathbf{s}_j)-
g^{-1}\dot{g}\dot{\mathbf{s}}_i
\label{eq:EM_w_s}\\
\ddot{h}&= \frac{1}{W}(\Pi -P)\eta 
\label{eq:EM_w_h}
\end{align} 
The tensor in equation \eqref{eq:EM_w_s} can be expanded to
\begin{equation}
g^{-1}\dot{g}=(h^T h)^{-1}(\dot{h}^T h + h^T \dot{h})
\end{equation}
Furthermore we have the following relation:
\begin{equation}
\eta=\left\{\mathbf{b}\times\mathbf{c},\mathbf{c}\times\mathbf{a},\mathbf{a}\times\mathbf{b}\right\}
\end{equation}

Since the forces are velocity-dependent, the time integration should be performed by an advanced integration scheme to correctly find the numerical solution of the equations of motion \eqref{eq:EM_w_s} and \eqref{eq:EM_w_h} (for example the symplectic Runge-Kutta method or a predictor-corrector scheme). However, we are only interested in escaping the local minimum and we only perform short MD runs with few oscillations in the potential energy. Therefore, the long-time conservation of the total energy is not an issue and we can simply use finite differences to discretize the equations of motion, equivalent to the Verlet algorithm.

If we define the following generalized forces
\begin{align}
\mathbf{f}_i&:=-\sum_{i\neq j}\frac{1}{r_{ij}}\frac{\partial \phi(r_{ij})}{\partial{r_{ij}}}(\mathbf{s}_i-\mathbf{s}_j)\\
F&:=(\Pi-P)\eta 
\end{align}
and use finite difference formulae for \mbox{$\ddot{\mathbf{s}}_i$}, \mbox{$\dot{\mathbf{s}}_i$} and $\ddot{h}$, the time evolution of the atomic positions and the cell vectors can be computed iteratively according to
\begin {align}
\label{eq:evo_pos}
\mathbf{s}_i(t+\Delta t)&=\mathbf{s}_i(t)+\Delta t \mathbf{v}_i(t)+\\
\nonumber &\qquad \Delta t^2\left( \frac{1}{m_i} \mathbf{f}_i(t)-g^{-1}(t)\dot{g}(t)\mathbf{v}_i(t)\right)\\
h(t+\Delta t)&=h(t)+\Delta t V(t) + \Delta t^2 \frac{1}{W}F(t)
\end{align}
Here we introduced the dummy variables \mbox{$\mathbf{v}_i(t):=\frac{\mathbf{s}_i(t)-\mathbf{s}_i(t-\Delta t)}{\Delta t}$} and \mbox{$V(t):=\dot{h}(t)=\frac{h(t)-h(t-\Delta t)}{\Delta t}$}, which can be considered as velocities of the corresponding variables.

\subsection{Softening and optimizing cell parameters}

\textit{Softening}, a method of biasing the initial velocities for the MD simulation of the escape step, is used to increase the efficiency of MH.~\cite{schoenborn_2009} For the crystal structure prediction MH the velocity vector consists not only of atomic velocities, but also of the cell velocities. First, a random velocity direction with Gaussian distributed magnitudes is chosen. The initial velocity amplitude is chosen that the kinetic energy is small, allowing only low barriers to be crossed during the MD escape. In chemistry, the Bell-Evans-Polanyi~\cite{jensen_1998} principle states that strongly exothermic reactions have a low activation energy. Reactant and product are in fact neighboring local minima on the potential energy surface and the chemical reaction is a transition of an energy barrier connecting the two minima. Hence, the Bell-Evans-Polanyi principle can be generalized to any transitions between local minima on the potential energy surface during MD simulations. Recently, Roy \textit{et al.} have shown that, on average, crossing low energy barriers along MD trajectories will lead into the basin of attraction of lower energy local minima than crossing high energy barriers.~\cite{roy_2008} Furthermore, low energy barriers are generally connected to low frequency eigenmodes of local minima.~\cite{sicher_2010} These properties can be readily extended to the configurational enthalpy. Therefore, the probability of finding low enthalpy configurations can be expected to increase when the direction of the initial velocity vector of a MD run points toward a direction with low curvature. Hence, in a second step, the velocity vector from the first step is rotated such that it is oriented along soft mode directions of the current minimum. The rotation procedure is performed by iteratively minimizing the energy along the escape direction at a constant distance from the local minimum.~\cite{schoenborn_2009} However, over-biasing the velocities is not favorable since the random and therefore ergodic character of the escape step should be retained in order not to arrive at a deterministic process. In fact, \textit{softening} is mainly applied to eliminate components of the velocity on hard modes.

If a quasi-Newton method (like the Broyden-Fletcher-Goldfarb-Shanno method) is used to perform local geometry optimization within the MH method an alternative \textit{softening} procedure is possible. In a quasi-Newton method an approximate Hessian matrix is continuously updated during a geometry relaxation. Before performing the MD escape step the approximate Hessian matrix from the previous relaxation step is diagonalized and a small number of low frequency eigenvectors are extracted. A randomized superposition of these eigenvectors are then used to provide the initial, soft velocity vector for the following MD trajectory. If feasible, the Hessian matrix can also be computed analytically to evaluate the soft mode directions.

\textit{Softening} has been successfully used in previous applications of MH.~\cite{schoenborn_2009,roy_2009,bao_2009} In Table \ref{tab:BLJ_soften} the performance of the MH method with and without \textit{softening} is compared for a benchmark system, a BLJ mixture with type A and B atoms in a small cell, A$_8$B$_4$. It can be clearly seen that the curvature of the configurational enthalpy along the velocity direction is reduced by roughly one order of magnitude when \textit{softening} is used. The significant increase in the efficiency when \textit{softening} is applied can also be ascribed to improving the cell velocities, since the impact of cell parameters on the structure can be larger than the atomic coordinates. For cells with a small number of atoms this can be illustrated by a simple example. Consider a simulation cell containing one single atom at the origin. The lattice vectors are given by \mbox{$\mathbf{a}=\frac{a}{2}(\hat{y}+\hat{z}-\hat{x})$}, \mbox{$\mathbf{b}=\frac{a}{2}(\hat{z}+\hat{x}-\hat{y})$} and \mbox{$\mathbf{c}=\frac{a}{2}(\hat{x}+\hat{y}-\hat{z})$} (where the hats denote the unit vectors and $a$ is the lattice constant), which define a body-centered lattice. Assume the atomic coordinates $\mathbf{x}$ were fixed. Transforming the cell to \mbox{$\mathbf{a}^*=\frac{a}{2}(\hat{y}+\hat{z})$}, \mbox{$\mathbf{b}^*=\frac{a}{2}(\hat{z}+\hat{x})$} and \mbox{$\mathbf{c}^*=\frac{a}{2}(\hat{x}+\hat{y})$} will result in a face-centered cubic lattice, a totally different structure. However, when the cell parameters $\mathbf{a}$, $\mathbf{b}$ and $\mathbf{c}$ are fixed, there is no possibility to transform the atomic coordinates $\mathbf{x}$ to a system where a face-centered cubic lattice is obtained. Obviously, the impact of the cell parameters decreases with increasing number of atoms, which is equivalent to the limit where an infinitely large cell is used. Similarly, this principle is also used in methadynamics simulations where the cell parameters are chosen as the collective variables.~\cite{laio_2002} Methadynamics simulations were recently applied to predict phase transitions in silicon based on neural-network representation of the density functional theory potential energy surface, where the Gibbs free energy surface was explored at fixed pressure and temperature as a function of the cell parameters $h$.~\cite{behler_2008} The effect of \textit{softening} on the efficiency of the MH method is in general larger for small periodic cells with a low number of atoms than for isolated molecules of similar size. So, for non-periodic systems with the same number of atoms as our benchmark cell \textit{softening} does not give a significant performance improvement.

\begin{table}[h!]
  \caption{The impact of \textit{softening} on various quantities of the MH method is shown. Starting from 100 different random input configurations all runs were continued until the ground state structure was found, resulting in a success rate of 100\%. The first column shows the different degrees of freedom (DOF) that are taken into account during \textit{softening}. The second and third column show the median value of the curvature of the enthalpy $\widetilde{\kappa}$ along the initial MD velocity direction before ($\widetilde{\kappa}_b$) and after ($\widetilde{\kappa}_a$) \textit{softening}, respectively. The fourth column contains the median value of the number of visited minima $\widetilde{n}_{\mathrm{min}}$ before reaching the global minimum. A BLJ mixture with A$_{8}$B$_{4}$ atoms was used described by the modified Lennard-Jones potential as discussed in section~\ref{sec:BLJ}. The parameters were set to $\sigma_{AA}=1.50$, $\sigma_{AB}=2.25$, $\sigma_{BB}=3.00$ and $\epsilon_{AA}=1.00$, $\epsilon_{AB}=1.25$, $\epsilon_{BB}=1.00$, and a cutoff radius of $r^{cut}_{\alpha\beta}=2.5\sigma_{\alpha \beta}$ was used.\\
  $^\S$The number of \textit{softening} iterations was doubled.\\
  $^\dagger$Initial atomic velocities were set to zero before \textit{softening}. }
\begin{ruledtabular}
\begin{tabular} {l r r r }
Softening DOF & $\widetilde{\kappa}_b$ $(\epsilon_{AA}/\sigma_{AA}^2)$ & $\widetilde{\kappa}_a$ $(\epsilon_{AA}/\sigma_{AA}^2)$& $\widetilde{n}_{\mathrm{min}}$  \\
    \hline
    None                       & 806.22 & 806.22  &  21.0 \\
    Atoms                      & 809.55 & 105.35  &  16.0 \\  
    Atoms \& Cell              & 813.62 & 74.18   &  11.5 \\ 
    Atoms \& Cell$^\S$         & 829.77 & 52.33   &  11.0 \\
    Atoms \& Cell$^\dagger$    & 111.32 & 64.47   &  10.5 \\
    Atoms \& Cell$^{\dagger\S}$& 112.62 & 50.73   &   8.0 \\ 
\end{tabular}
\end{ruledtabular} 
\label{tab:BLJ_soften}
\end{table}

The fictitious cell mass $W$ is another adjustable parameter. Choosing a too large ratio for cell and atomic mass will result in a very stiff cell which will not adjust itself smoothly during the simulation. However, when this ratio is to small the cell can fluctuate violently resulting in a strong deformation of the cell within one step of the simulation. We have found that choosing the cell mass similar to the atomic masses is a reasonable choice.

During a MH simulation it can happen that the cell shape gets heavily distorted leading to small angles between the three lattice vectors, and thus resulting in a very flat cell. This behaviour is not desired since more periodic cells have to be taken into account when computing the potential energy and its derivatives. Furthermore, it makes it difficult to identify equivalent structures in different cells. Therefore, whenever necessary, a transformation of the cell vectors is performed to obtain shorter cell vectors (for details see Ref.~[\onlinecite{oganov_2008}]).

\subsection{Application on silicon with optimized performance by parallelization and lattice vector prediction}

As an application of the MH method, a silicon supercell with 64 atoms was investigated at zero pressure. Since the number of local minima increases exponentially with respect to the number of atoms in a system, finding the global minimum of a cell with many atoms is a challenging task. Furthermore, the silicon supercell with 64 atoms is a multi-funnel system with crystalline and amorphous structures and therefore presents an additional challenge for global optimization. For a silicon supercell with 64 atoms in a rigid box~\cite{goedecker_2004} the difficulty of finding the ground state has already been demonstrated. Several million minima had to be visited before finding the ground state, the well-known cubic diamond structure. To demonstrate the advantage of the variable cell shape MH, we revisit the same problem set. Using the environment-dependent interatomic potential (EDIP) for silicon~\cite{bazant_1996,bazant_1997,justo_1998} statistical data were collected for a set of 100 serial MH runs at zero pressure. Each run started with a highly random configuration in a distorted cell and was stopped as soon as a structure with ground state energy was found or 8000 distinct minima were accepted, a small number for such a large system.
Since some of the local minima are visited several times and only half of the new structures are accepted the search length corresponds to visiting at most some \mbox{50 000} minima. 
The success rate was found to be 45\%. Each successful run visited an average of some \mbox{13 000} local minima till finding a structure with ground state enthalpy, an improvement in performance by two orders of magnitude compared to the earlier results. However, since the EDIP potential has only first neighbor interaction, the cubic diamond structure and its polytypes (for example the hexagonal diamond (Lonsdaleite) structure) all give the same ground state enthalpy. Therefore, only 12 out of the 45 successful runs arrived at the cubic diamond structure. It needs to be emphasized that this is not a shortcoming of the MH method, but an effect of the short range character of the EDIP potential. Increasing the number of distinct and accepted minima to be found will lead to an increase in the success rate. So, when increasing the search length by a factor of 6 the success rate is almost doubled to 80\%.

\begin{figure}            
\setlength{\unitlength}{1cm}
\subfigure[]{\includegraphics[width=0.9\columnwidth,angle=0]{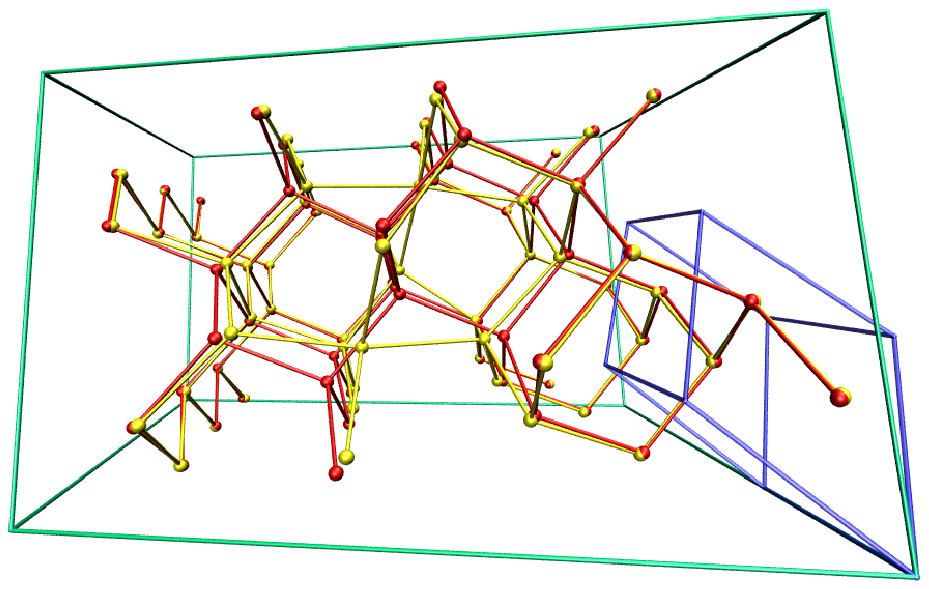}}  
\subfigure[]{\includegraphics[width=0.9\columnwidth,angle=0]{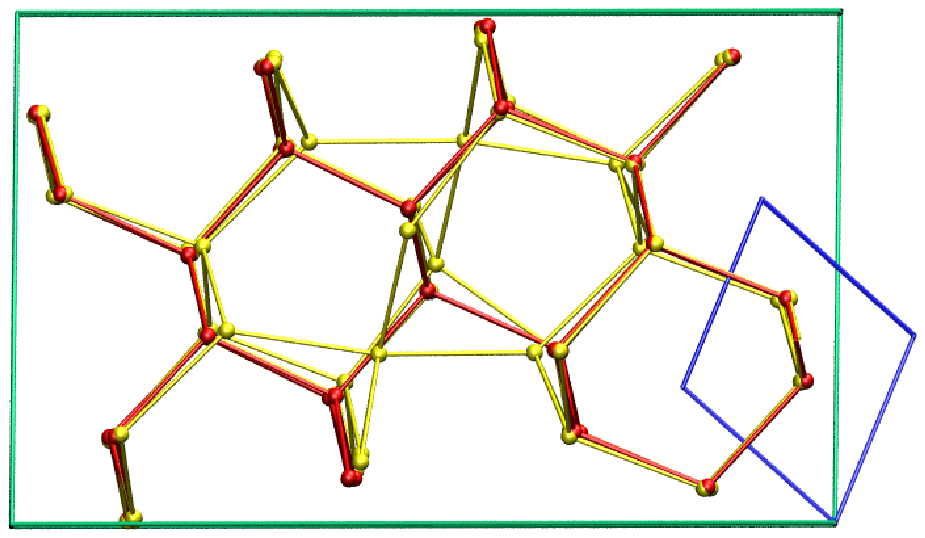}}  
\caption{The LVPS is illustrated for a Si$_{64}$ supercell in a perspective (a) and a orthographic (b) view. The yellow (light) and red (dark) spheres denote the positions of the atoms before and after applying LVPS, respectively. The resulting supercell consists of 66 atoms. The primitive cell found by the LVPS is shown in the right bottom corner by a blue parallelepipede.}
\label{fig:LVPS}
\end{figure}

Another approach to increase the success rate of the MH method is by parallelizing the MH runs. In our experience, the success rate of a serial MH simulation can depend heavily on the initial guess of the structure. Therefore, it can be advantageous to start multiple parallel simulations starting from different initial structures. We performed a simple statistical analysis to estimate the computational cost necessary if, instead of performing multiple serial runs with index $i$, a parallel run is started with $m$ processes of which each requires $l_i$ MH steps to find the ground state. In the parallel version all processes would be stopped as soon as one of the processes finds the ground state structure. The necessary number of minima to be visited is given by $n_{\mathrm{min}}^{m}=m \cdot  \mathrm{min}\{l_i\}$ where $i=1,\dots,m$. An average value $\overline{n}_{\mathrm{min}}^{m}$ should be computed from runs with different initial structures. Then, the optimal number of processes for the particular problem is $m=m_0$ resulting in the minimal value of $\overline{n}_{\mathrm{min}}^{m}$. Applied to the Si$_{64}$ system the computational cost can be reduced to half for $m_0=6$, $\overline{n}_{\mathrm{min}}^{6}\approx6200$

For non-periodic systems the identification of structures based on their energies is sufficiently accurate in most cases, but the enthalpy degeneracy of the polytypes within shortranged potentials requires some additional method to distinguish different structures. Since the most natural characteristic of a structure is its geometry we use a fingerprint function related to experimental diffraction patterns proposed by Valle \textit{et al.}.~\cite{Valle2008} A continuous one-dimensional function is defined by summing up weighted Gaussian functions centered at all relative atomic distances. To reduce numerical errors the fingerprint function is discretized into $m$ bins, leading to a vector of size $m$ uniquely related to the structure. By using the angle between two fingerprint vectors a cosine distance is defined which then can be used to determine the similarities between structures.

\begin{figure}[bh!]            
\setlength{\unitlength}{1cm}
\includegraphics[width=0.9\columnwidth]{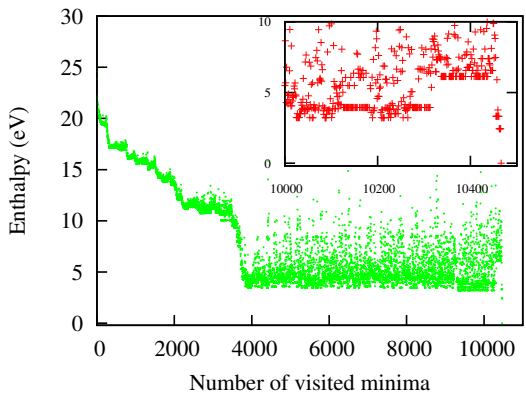}  
\caption{The enthalpies of the local minima visited during a MH simulation of a Si$_{64}$ supercell at zero pressure, starting from a random input configuration. The enthalpies were shifted such that the ground state enthalpy is zero. The inset shows clearly the crossing of the potential barrier before arriving at the ground state structure.}
\label{fig:MHM_Energy}
\end{figure}

Frequently, when a global optimization run does not reach the global minimum for a long time, the simulation is stuck in an enthalpy funnel with barriers hard to overcome. In most cases of our simulation of \mbox{Si$_{64}$} at zero pressure these funnels are determined by lattice vectors which can fit a cubic diamond structure with 62 or 66 silicon atoms, but not exactly the given 64 atoms. In these cases, a diamond structure (or one of its polytypes) fit into the cell perfectly with exception of some defective subregions where 2 Si atoms are either missing or are redundant. For these cases we developed a lattice vector prediciton scheme (LVPS) to modify the simulation cell by adding or removing atoms such that a perfect crystalline structure is recreated. We define a three dimensional scalar function $f(\mathbf{r})$ by summing up Gaussian functions with a width $\sigma$ on each atomic position $\mathbf{r}_i$

\begin{equation}
 f(\mathbf{r})=\sum_{i=1}^{N}\frac{1}{\left(2\pi\sigma^2\right)^{(3/2)}}\exp\left(-\frac{(\mathbf{r}_i-\mathbf{r})^\intercal(\mathbf{r}_i-\mathbf{r})}{2\sigma^2}\right)
\end{equation}

The Gaussian width should be small enough that the overlap between neighboring atoms is vanishingly small. The autocorrelation function is defined as
\begin{equation}
h(\mathbf{r})=\int_{-\infty}^\infty f^*(\mathbf{\boldsymbol\tau})f(\mathbf{r}+\mathbf{\boldsymbol\tau})d\mathbf{\boldsymbol\tau}
\end{equation}

In Fourier space this convolution is simply a multiplication of the individual fourier transformed functions of $f^*$ and $f$, and according to the Wiener-Chinchin theorem
\begin{equation}
H=\mathcal{F}(h)=(\mathcal{F}(f))^*\cdot\mathcal{F}(f)=|\mathcal{F}(f) |^2
\end{equation}
where $\mathcal{F}$ denotes the Fourier transform and $^*$ denotes the complex conjugate. A transformation back to real space results in the desired autocorrelation function. The autocorrelation function is then scaled such that the peak at the origin is 1, and periodic boundary conditions are applied. The three peaks closest to the origin with an amplitude of more than 0.5 spanning a parallelepipede with a non-vanishing volume give the vectors of a primitive unit cell. Using this cell to span the whole simulation cell, the most common basis is identified and the crystalline structure is reproduced (see Fig.~\ref{fig:LVPS}). Using LVPS to identify ground state structures the success rate was further increased to almost 95\%. The LVPS can be further used to predict the correct system size when the number of atoms in the crystal basis is not known in advance.

Fig.~\ref{fig:MHM_Energy} shows a typical progress of a MH run for a Si$_{64}$ supercell at zero pressure. First, starting from a random structure, the enthalpy decreases as new parts of the configurational enthalpy surface are explored. After visiting some 4000 local minima, the system is caught for some time in a deep funnel. This funnel corresponds exactly to the case discussed above and the perfect crystal could be completed applying the LVPS. However, after visiting slightly more than \mbox{10 200} minima a barrier is crossed and the system finally reaches the ground state structure.

The EDIP potential proved to be somewhat inaccurate when predicting the enthalpies of structures at a pressure of 16 GPa. Both the well-known $\beta$-tin structure~\cite{jamieson_1963,hu_1984,olijnyk_1984} (Si-II) and the Imma structure~\cite{mcmahon_1993} (Si-XI) were found to be metastable in EDIP. The simple hexagonal phase~\cite{olijnyk_1984} (Si-VII) is not even a local minimum on the enthalpy surface and relaxes to the simple cubic structure. Instead, a structure with shifted layers of cubic elements where each atom is fourfold coordinated with a bond lenght \mbox{$\approx2.4$ \AA} was predicted as the ground state with an enthalpy of $-2.923$~eV and a volume of 14.378~\AA$^3$ per atom, a structure not in the fitting database used for EDIP. Finding this unexpected crystalline ground state shows the predictive power of the MH method for unknown structures. However, the novel structure was found to be unstable within density functional theory calculations. The second lowest crystalline structure was found to be the bct-5 structure~\cite{boyer_1991} with an enthalpy of $-2.865$~eV and a volume of 15.264~\AA$^3$ per atom.

\begin{figure}[bh!]            
\setlength{\unitlength}{1cm}
\includegraphics[width=0.9\columnwidth]{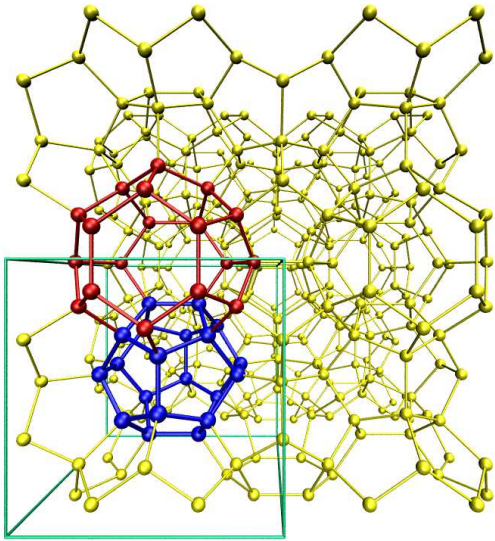}  
\caption{The type-I silicon clathrate is shown with the unit cell represented by the green box. Two types of cages are used to compose the structure, a small pentagonal dodecahedron (blue) and a larger hexagonal truncated trapezohedron (red).~\cite{weaire_1994}}
\label{fig:Si46}
\end{figure}

As a next application we studied the global optimization scheme on a Si$_{46}$ supercell at zero pressure with volume constraints which was realized by introducing a harmonic energy term $E_{vol}$ (second term in equation~\eqref{eq:EDIP_vol}) in addition to the standard EDIP potential energy: 
\begin{equation}
E_{tot}=E_{EDIP}+k\cdot(\det(h)-\Omega_0)^2
\label{eq:EDIP_vol}
\end{equation}
where $k$ is a small scalar value and $\Omega_0$ is the target volume. The additional gradient on the cell vectors is given by
\begin{equation}
\frac{\partial E_{vol}}{\partial h_{\alpha\beta}}=2k\cdot(\det(h)-\Omega_0)\det(h)h_{\beta\alpha}^{-1}
\end{equation}
The Si$_{46}$ was chosen since it can form the unit cell of a type-I Si$_{46}$ clathrate structure~\cite{adams_1994} (see Fig.~\ref{fig:Si46}). Although clathrates have fourfold coordinated structures their overall geometry differ significantly from the cubic diamond structure. Composed from polyhedral building blocks, a major part of the unit cells remains void, resulting in porous crystals. Therefore, the type-I Si$_{46}$ clathrate within the EDIP potential has a large volume per atom ratio of 22.852 \AA$^3$.

Starting from random input positions 20 MH runs were started using the equilibrium volume of the type-I Si$_{46}$ clathrate as the target $\Omega_0$ and $k=2.0$, each process visiting some \mbox{1.5 Mio.} minima. Only one run was able to find the clathrate ground state unit cell after visiting roughly \mbox{150 000} minima. Nevertheless, finding the clathrate unit cell at all is an encouraging result since this particular system is a big challenge in global optimization for the following reasons. First, there is only one unit cell corresponding to the ground state and it consists of a huge basis of 46 atoms with a complex structure. Second, there are two main funnels that compete during the search for the ground state. On one hand, the system prefers to crystallize to the cubic diamond structure, but the void areas with the dangling bonds are energetically not favorable. On the other hand there is a tendency of forming spherical cage-structures in a porous crystal, but it is seldom possible to obtain tetrahedral bond angles. These are two fundamentally different structures and are separated by a very high potential barrier hard to overcome, hence starting from many different random input guesses is crucial for a successful run.

\begin{figure}[bh!]            
\setlength{\unitlength}{1cm}
\subfigure[]{\includegraphics[width=0.45\columnwidth]{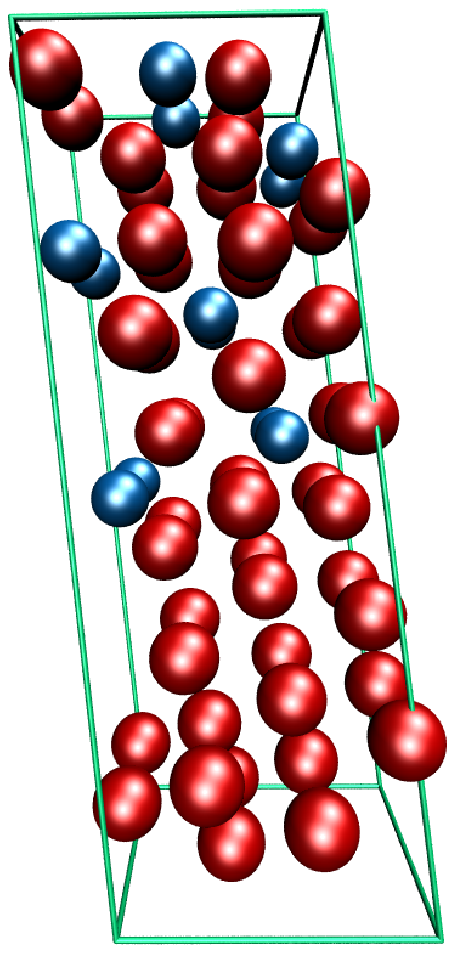}}  
\subfigure[]{\includegraphics[width=0.45\columnwidth]{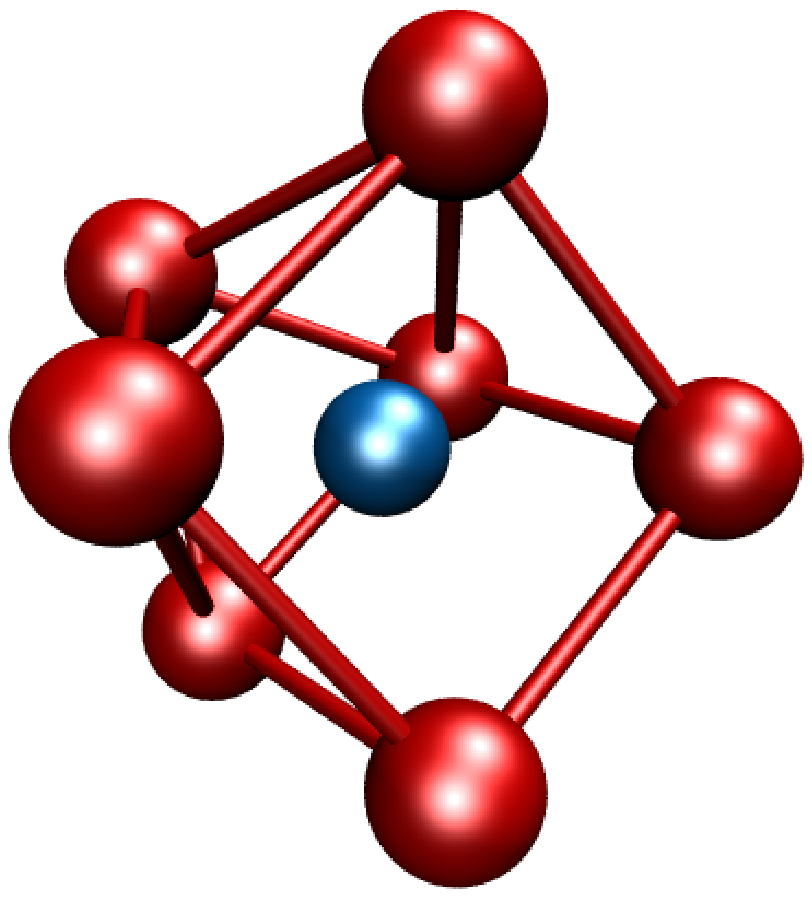}}
\caption{(a) A cell of the putative ground state of a BLJ crystal consisting of 48 type A and 12 type B atoms. Type A atoms are denoted by red (large) spheres, type B atoms are blue (small). The detailed structure of the embedded type B atoms is shown in subfigure (b). The 7 type A atoms form a monocapped trigonal prism.~\cite{fernandez_2004} The symmetry group of this isolated molecule is $C_{2\nu}$}
\label{fig:LJ60}
\end{figure}

\subsection{Application on binary Lennard-Jones mixtures}
\label{sec:BLJ}
As a last application we reinvestigated a much-studied BLJ mixture at zero pressure, a system widely accepted as benchmark systems. Lately, these mixtures have been studied by Middelton \textit{et al.}.~\cite{middleton_2001} They found that ordered crystalline phases are energetically favored, contrary to earlier results indicating a preference for glassy, amorphous structures. The putative ground state structures found in this previous work are available on the Cambridge Cluster Database.~\cite{CCD} We studied a supercell with 60 atoms consisting of 80\% type A and 20\% type B components. In our calculations we use a small modification of the well-known Lennard-Jones potential~\cite{jones_1924} (also used in Ref.~[\onlinecite{middleton_2001}]), truncated and shifted using a quadratic function such that both the energy and the first derivative are continuous at the cutoff distance.~\cite{stoddard_1973} The functional form of the pair potential is given in equation~\eqref{eq:LJ} where the subindices $\alpha$ and $\beta$ denote the atom types A and B, $r_{\alpha\beta}$ is the interatomic distance, $\epsilon_{\alpha\beta}$ is the potential well depth (not to be confused with the strain tensor) and $\sigma_{\alpha\beta}$ corresponds to the distance where the potential vanishes. The potential is zero when the radial distance is larger than the cutoff $r^{cut}_{\alpha,\beta}$. All parameters are identical to the ones used in Ref.~[\onlinecite{middleton_2001}], namely: $\sigma_{AA}=1.00$, $\sigma_{AB}=0.80$, $\sigma_{BB}=0.88$, $\epsilon_{AA}=1.00$, $\epsilon_{AB}=1.50$, $\epsilon_{BB}=0.50$, and a cutoff radius of $r^{cut}_{\alpha,\beta}=2.5\sigma_{\alpha \beta}$. All energies and enthalpies are given in units of $\epsilon_{AA}$.

\begin{multline}
\phi_{\alpha\beta} =  4 \epsilon_{\alpha\beta} \left\{ \left[ \left( \frac{\sigma_{\alpha\beta}}{r_{\alpha\beta}} \right) ^{12} - \left( \frac{\sigma_{\alpha\beta}}{r_{\alpha\beta}}\right)^6 \right]\right.\\
+\left[ 6\left(\frac{\sigma_{\alpha\beta}}{r^{cut}_{\alpha,\beta}}\right)^{12}- 3\left(\frac{\sigma_{\alpha\beta}}{r^{cut}_{\alpha,\beta}}\right)^6 \right]\left( \frac{r_{\alpha\beta}}{r^{cut}_{\alpha,\beta}}\right)^2\\
-\left. 7\left(\frac{\sigma_{\alpha\beta}}{r^{cut}_{\alpha,\beta}}\right)^{12}+4\left(\frac{\sigma_{\alpha\beta}}{r^{cut}_{\alpha,\beta}}\right)^6 \right\} 
 \label{eq:LJ}
\end{multline}

We found several thousand structures with enthalpies lower than the -7.08 $\epsilon_{AA}$ per atom of the crystal previously found in Ref.~[\onlinecite{middleton_2001}]. The putative ground state was found to have an enthalpy of -7.49 $\epsilon_{AA}$ per atom. In fact, this value is even lower than any other enthalpy of the larger supercells which were investigated by Middelton \textit{et al.}. The periodic cell of the ground state structure is shown in Fig.~\ref{fig:LJ60}~(a). It consists of two different regions similar to a phase separation. The lower section consists purely of type A atoms in a slightly distorted hexagonal closed packed structure. The upper half of the cell consists of a mixed phase where type B atoms are embedded in cages consisting of 7 A-type atoms. A detailed illustration of this behaviour is shown in Fig.~\ref{fig:LJ60}(b).

\section{Conclusions}

We have generalized the MH method to periodic systems using variable cell shape MD for the escape step. To cross low-enthalpy barriers more quickly along the MD trajectory the velocities are chosen to point along soft mode directions, taking also into account the lattice parameters. In several applications we show that low enthalpy structures can be predicted and ground state configurations can be found efficiently. For silicon systems the ground state structure at zero pressure could be found orders of magnitude faster compared to the initial MH method with fixed cell shape. Using parallel runs and the LVPS, the success rate can be significantly increased. The LVPS also allows to find unknown crystalline structures without knowing the exact number of atoms of the unit cell in advance.
For the Si$_{46}$ supercell the type-I clathrate structure was found using volume constraints. Small BLJ mixture cells were used to demonstrate the effect of \textit{softening} on the optimization performance. For larger BLJ mixtures a much-studied composition was reinvestigated and new putative ground state structures were found. Overall, the periodic MH method has shown to be a promising approach in structure prediction.

\section{Acknowledgments}

We thank \mbox{A. R. Oganov} and \mbox{T. J. Lenosky} for interesting expert discussions. Financial support provided by the Swiss National Science Foundation are greatfully acknowledged. Computational resources were provided by the Swiss National Supercomputing Center (CSCS) in Manno.

\end{document}